\documentclass[aps,twocolumn, pra]{revtex4}
\usepackage{graphicx}
\usepackage{epsfig}
\usepackage{newlfont}
\usepackage{amssymb}
\usepackage{amsfonts}

\usepackage{amsmath}
\usepackage{mathbbol}
\usepackage{graphicx}
\usepackage{bm}
\usepackage{natbib}

\begin{document}
\title{Particle Counting Statistics of Time and Space Dependent Fields}

\author{Sibylle Braungardt\(^1\), Mirta Rodr\' iguez\(^2\) , Roy J. Glauber\(^3\), and Maciej
Lewenstein\(^{1,4}\)} \affiliation{\(^1\)ICFO-Institut de
Ci\`encies Fot\`oniques, Mediterranean Technology Park,
08860 Castelldefels (Barcelona), Spain\\
\(^2\)Instituto de Estructura de la Materia, CSIC, C/Serrano 121 28006 Madrid, Spain \\
\(^3\)Lyman Laboratory, Physics Department, Harvard University, 02138 Cambridge, MA, U.S.A. \\
\(^4\)ICREA-Instituci\`o Catala de Ricerca i Estudis Avan{\c
c}ats, 08010 Barcelona, Spain}

\begin{abstract}
The counting statistics give insight into the properties of
quantum states of light and other quantum states of matter such as
ultracold atoms or electrons. The theoretical description of
photon counting was derived in the 1960s and was extended to
massive particles more recently. Typically, the interaction
between each particle and the detector is assumed to be limited to
short time intervals, and the probability of counting particles in
one interval is independent of the measurements in previous
intervals. There has been some effort to describe particle
counting as a continuous measurement, where the detector and the
field to be counted interact continuously. However, no general
formula applicable to any time and space dependent field has been
derived so far. In our work, we derive a fully time and space
dependent description of the counting process for linear quantum
many-body systems, taking into account the back-action of the
detector on the field. We apply our formalism to an expanding
Bose-Einstein condensate of ultracold atoms, and show that it
describes the process correctly, whereas the standard approach
gives unphysical results in some limits. The example illustrates
that in certain situations, the back-action of the detector cannot
be neglected and has to be included in the description.
 \end{abstract}
\maketitle

\def\com#1{{\tt [\hskip.5cm #1 \hskip.5cm ]}}

\def\bra#1{\langle#1|} \def\ket#1{|#1\rangle}
\def \av#1{\langle #1\rangle}
\section{Introduction}
The counting statistics of photons have been used to identify the
quantum state of light since the beginnings of quantum optics.
Despite early attempts \cite{Mandel1958,Mandel1959,Mandel1963} to
use a semiclassical approach, the theoretical description of the
counting process requires a full quantum mechanical description of
the electromagnetical field and the photocounter. This task was
first achieved by Glauber \cite{Glauber1963a, Glauber1963b,
Glauber1963c,Glauber1965}, for an analogous approach see
\cite{Kelley1964}, and will be referred to as the quantum counting
formalism in the following. The formalism uses a full
quantum mechanical description of the interaction between the
incoming light and the atoms in the detector. However, consecutive
detection events are treated independently, such that the effect
of absorption of photons at the detector is neglected.

Even though the quantum counting formalism is being applied in a
wide range of situations, its limits have been pointed out and
discussed in the last decades. For example, for long detection
times, it may lead to negative probabilities or expectation values
exceeding the total number of photons. Such limitations were first
pointed out by Mollow \cite{Mollow1968} and several authors
derived a formula taking into account the back-action of the
detector on the field by considering the evolution of the system
composed of the detector and the field
\cite{Mollow1968,Scully1969,Selloni1978}. In 1981, Srinivas and
Davies \cite{Srinivas1981} provide a systematic description of
photon counting as a continuous measurement process, where the
detector continuously interacts with the light field. They derive
their formalism based on a one-count operator and a no-count
operator that determine the time evolution of the combined system
of the detector and the light field. Whereas these operators are
postulated in the model by Srinivas and Davies (SD model), Imoto
and coworkers derived a microscopic theory of the continuous
measurement of the photon number \cite{Imoto1990}. They derive the
no-count and one-count operators taking into account the
interaction of the photons with the detector using the
Jaynes-Cumming Hamiltonian for the field of a two-level atom.
Photon counting including the back action of the detector has also
been treated in \cite{Dodonov2007,Hayrynen2010a}.

Contrary to the quantum counting formalism, the SD model predicts
that a substraction of one photon from a thermal state can
increase the expectation value of the number of photons
\cite{Ueda1990}. Recently, this has been confirmed in an
experiment where the action of the annihilation operator on
different states of light has been implemented and measured
\cite{Parigi2007}.  As discussed in  \cite{Hayrynen2009} this is
an experimental confirmation of the SD model.

The previous discussion suggests, that the SD model is a valid
description for the photon counting process based on continuous
measurements. However, a closed formula is
derived only for the case of a single-mode field and the application
of the formula to real experimental situation is therefore limited
\cite{Mandel1981,Fleischhauer1991}. In 1987, Chmara derived a
general formula for the photon counting distribution for a
multimode field \cite{Chmara1987} by applying the photon counting
approach by SD to an open system. While the formula is in
principle applicable to a wide class of systems, to our knowledge,
no practical case where the time dependent intensities have been
calculated has been reported.

In our work, we study the role of the back-action of the detector
when counting massive particles, however, the results are valid
also for photons. Previously we have shown that the
particle counting statistics can be used as a detection method for
many-body systems of ultracold atoms \cite{Braungardt2008,
Braungardt2011a, Braungardt2011b}. In this work we derive a
general formula that describes the counting process including the
back-action of the detector on the field. In contrast to the
earlier descriptions of the counting process stated above, we give
a fully time- and space dependent description of the process.
We illustrate the importance of the back-action of the detector on
the field by applying our formula to the detection of an expanding
Bose Einstein condensate (BEC) by particle counting. We compare
our formalism to the quantum counting formalism and to an approximate
solution obtained by the Born approximation. We show that the
approximate solutions, although more accurate than the quantum counting formalism, still fail to describe the counting process
correctly.

The paper is organized as follows: In section \ref{sec:I} we
review the quantum counting formalism and the SD model. In section
\ref{sec-time-dep} we consider a fully time- and space dependent
description of the counting process and derive a formula for the
counting distribution which is extended for two detectors in Sect.
\ref{sec:2d}. We illustrate the space and time dependent counting
process by considering the counting statistics of a freely
expanding Bose Einstein condensate in Sect. \ref{sec-system-td}.
We compare our time- and space-dependent counting formalism to the
quantum counting formalism and show that it some limits, the last
one leads to a divergent intensity at the detector. We also
analyze the effect of the absorptive part in the modified field
operator by comparing the exact solution to other approximative
methods. We summarize our results in Sect.\ref{sec:summ}.

\section{Photon Counting: Standard approach and continuous measurement approach\label{sec:I}}
In this section, we review the two main approaches to photon
counting: The quantum counting formalism derived in the works of
Glauber \cite{Glauber1963a, Glauber1963b,
Glauber1963c,Glauber1965} and the continuous measurement approach
derived by Srinivas and Davies \cite{Srinivas1981}.
\subsection{The Quantum Counting Formalism}
The derivation of the quantum counting formalism is based on the
description of the quantum mechanical interaction of the photons
with the atoms in the detector. The approach uses perturbation
theory to describe the interaction for short intervals of time.
The counting distribution for the full detection time $\tau$ is
obtained by dividing it into small subintervals $\Delta\tau$ and
treating the measurement in the full interval as a number of
successive independent measurements in each interval. The approach
thus describes a sequence of measurements, where the field evolves
as in the absence of the detector. The method is based on the
assumption, that the detection in one subinterval is independent
of the detection in the previous subintervals. For the case of a
light beam falling on a photo detector, it is argued \cite[p.
723]{Mandel1995} that each element of the optical field interacts
with the detector only briefly. The response time of the detector
is short and the energy of the electron state is well defined
after an interaction time $\Delta\tau$. For such a system where
the unabsorbed photons escape, there is no need to consider the
measurement back-action. The resulting equation for the counting
distribution reads
\begin{equation}
p(m,t,\tau)=\langle\mathcal{T}:\frac{(\mathcal{I}(t))^m
e^{-\mathcal{I}(t)}}{m!}:\rangle.\label{eq-quantum-mandel}
\end{equation}
Here $\mathcal{T}$ and $::$ stand for time and normal ordering,
respectively. The intensity operator $\mathcal{I}(t)$ is defined
in terms of the positive and negative frequency parts
$\hat{E}^\pm(\textbf{r},t)$ of the field operators by
$\mathcal{I}(t)=\epsilon\int_t^{t+\tau}dt'\int_\Omega
\hat{E}^-(\textbf{r},t)\hat{E}^+(\textbf{r},t)d\textbf{r}$, where
the spatial integral runs over the detector area $\Omega$ and
$\epsilon$ denotes the efficiency of the detector. The normal
ordering reflects the fact that the photons are annihilated at the
detector. For a single mode field, the formula reads
\begin{equation}
p(m,\tau)=\sum_{n=m}^\infty\binom{n}{m}(\epsilon
\tau)^m(\exp(-\epsilon \tau))^{n-m}\langle n|\rho
|n\rangle.\label{eq-mandel-free}
\end{equation}
where $\rho$ is the density matrix  of the measured system.
\subsection{The Back-action of the Detector on the Field}
In  \cite{Srinivas1981} Srinivas and Davies developed an approach
to photon counting based on continuous measurements over an
extended period of time. Their work was motivated by the fact that
the quantum counting formula Eq. (\ref{eq-quantum-mandel})
exhibits inconsistencies in the limit of long detection times,
such as negative probabilities or mean particle numbers that
exceed the total number of particles. For a single mode field the
counting probability distribution is given by
\begin{equation}
p(m,t,\tau)=\sum_{n=m}^\infty\binom{n}{m}(1-e^{-\epsilon
\tau})^m(\exp(-\epsilon \tau))^{n-m}\langle n|\rho
|n\rangle.\label{eq-mollow}
\end{equation}
This expression is formally equivalent to the quantum counting
formula Eq. (\ref{eq-mandel-free}) with the term $\epsilon \tau$
substituted by $1-e^{-\epsilon \tau}$. For $\epsilon \tau\ll 1$,
the two formulas are equivalent but the SD counting distribution
in Eq. (\ref{eq-mollow}) does not suffer from the inconsistencies
as the quantum counting formalism. The mean number of photons that
are registered is bounded for $\tau\rightarrow\infty$ and no
negative probabilities occur. However, the applicability of the
formalism derived by Srinivas and Davies is limited to a one-mode
field and therefore fails to describe most experimental
situations.

Summarizing, the quantum counting formalism Eq.(\ref{eq-mandel-free}) becomes meaningless if
$\epsilon \tau \gg 1$, when it can result in negative
probabilities or unlimited mean number of counted photons as $\tau$
tends to infinity.  Most experimentally relevant
situations of photon counting are not in this limit and thus are well
described by the quantum counting formalism. In our work we are
interested in the counting statistics of many-body systems of cold
atoms where one can control the space and time dependence. In the following section, we derive a formula for the
counting distribution that takes into account the back-action of
the detector on the field. It generalizes the formalism developed by
Srinivas and Davies to time- and space dependent fields.

\section{Particle Counting of Time and Space Dependent
Fields}\label{sec-time-dep} The time- and
space-dependent counting process can be described by a master
equation that describes the interaction between the detector and
the detected field $\Psi(\textbf{r},t)$.
In typical experimental situations, the interaction between the detector and the
field is restricted to a given volume, such as the surface of a photodetector or a microchannel plate.
We define the function $\Omega(\textbf{r})$ to describe the spatial configuration of the detector.  The master equation then reads
\begin{eqnarray}
\dot{\rho}(t)=\epsilon \int d\textbf{r}
\Omega(\textbf{r})\Psi(\textbf{r},t)\rho
\Psi^\dag(\textbf{r},t)\nonumber\\-\frac{\epsilon}{2}\int
d\textbf{r} \Omega(\textbf{r})[\Psi(\textbf{r},t)^\dag
\Psi(\textbf{r},t) \rho-\rho \Psi(\textbf{r},t)^\dag
\Psi(\textbf{r},t)],\label{eq-master-Psi}
\end{eqnarray}
where $\rho(\textbf{r},t)$ denotes the density matrix of the
system. The first term on the right-hand side of eq.
(\ref{eq-master-Psi}) corresponds to the number of quantum jumps
in the detector volume whereas the second term represents the
damping of the field due to the absorption at the
detector. \\
In order to solve the master equation (\ref{eq-master-Psi}) we
first perform the transformation
$\rho(t)=F(t)\tilde{\rho}(t)F^\dag(t)$ and define
\begin{equation}
\tilde{\Psi}(\textbf{r},t)=F^{-1}(t)\Psi(\textbf{r},t)F(t),\label{eq-def-psitilde}
\end{equation}
where the operator $F(t)$ is defined as
\begin{eqnarray}
F(t)=\mathcal{T}e^{-\epsilon/2\int_0^t dt' \int
d\textbf{r}'\Omega(\textbf{r}')\Psi^\dag(\textbf{r}',t')\Psi(\textbf{r}',t')}.
\label{eq-def-F(t)}
\end{eqnarray}
Here, the term $\mathcal{T}$ on the left side of the operator
denotes time ordering, whereas it denotes opposite time ordering
on the right side of the operator. We use the relation
$e^{\gamma A}Be^{-\gamma
A}=B+\gamma[A,B]+\frac{\gamma^2}{2!}[A,[A,B]]+...$
 and the
commutation relations for linear fields,
  $[\Psi(\textbf{r},t),\Psi^\dag(\textbf{r}',t)]=\delta(\textbf{r},\textbf{r}')$,
 $[\Psi(\textbf{r},t),\Psi^\dag(\textbf{r}',t')]=G_0(\textbf{r},\textbf{r}'|t,t')$, $
 [\Psi(\textbf{r},t),\Psi(\textbf{r}',t')]=0$,
where $G_0({\textbf{r}},{\textbf{r}'}|t,t')=\bra{\textbf{r}}
e^{-i(t-t')/\hbar H_0}\ket{\textbf{r}'} $ is the propagator for
the time evolution for the Hamiltonian $H_0$ that describes the
evolution of the field without detection.

The evolution of the rotated density matrix fulfills the equation
\begin{equation}
\dot{\tilde{\rho}}(\textbf{r},t)=\epsilon\int
d\textbf{r}'\Omega(\textbf{r}')\tilde{\Psi}(\textbf{r},t)\tilde{\rho}\tilde{\Psi}^\dag(\textbf{r},t).\label{eq-dot_rhotilde}
\end{equation}
A master equation of this kind, for non-rotated $\Psi$ and $\rho$
instead of $\tilde{\Psi}$ and $\tilde{\rho}$ leads to the quantum
counting formalism described by eq. (\ref{eq-quantum-mandel}). Eq.
(\ref{eq-dot_rhotilde}) can be solved using perturbation theory
(note that $\tilde{\rho}(0)=\rho(0))$, such that
\begin{widetext}
\begin{eqnarray}
&\tilde{\rho}(t)=\rho(0)+\epsilon\int_0^tdt'\int d\textbf{r}'\Omega(\textbf{r}') \tilde{\Psi}(\textbf{r}',t')\rho(0)\tilde{\Psi}^\dag(x',t')\\
&+\epsilon^2\int_0^t dt'\int_0^{t'}dt''\int d\textbf{r}'\int
d\textbf{r}''\Omega(\textbf{r}')\Omega(\textbf{r}'')\tilde{\Psi}(x',t')\tilde{\Psi}(x'',t'')\rho(0)\tilde{\Psi}^\dag(x'',t'')\tilde{\Psi}^\dag(x',t')+...\nonumber
\label{eq-rhotilde}
\end{eqnarray}
\end{widetext}
We use the cyclic properties of the trace to calculate the
probability $p(m,\tau)$ of finding $m$ particles within the
detector opening time $\tau$ and from the $m$th order term in the
expansion in eq. \ref{eq-rhotilde} we get
\begin{eqnarray}
p(m,\tau)=\langle\epsilon^m\int_0^\tau dt'\int_0^{t'}dt''..\int
d\textbf{r}'\int
d\textbf{r}''\Omega(\textbf{r}')\Omega(\textbf{r}'')..\\\nonumber
\tilde{\Psi}^\dag(x',t')\tilde{\Psi}^\dag(x'',t'')..F^\dag(\tau)F(\tau)..\tilde{\Psi}(x'',t'')\tilde{\Psi}(x'',t')\rangle.\label{eq-p-rhotilde}
\label{eq-p-not-normally}
\end{eqnarray}
We rewrite this expression as a normal ordered expression with
respect to the modified operators $\tilde{\Psi}$, which is also
normal ordered with respect to the operators $\Psi$, as they are
related by a linear transformation. Taking into account the
normalization of the counting distribution, we obtain
\begin{eqnarray} p(m,\tau)
=\langle\mathcal{T}:\frac{(\mathcal{I}(\tau))^m}{m!}e^{-\mathcal{I}(\tau)}:\rangle,\label{eq-p-time-dependent}
\end{eqnarray}
where the intensity at the detector is given by
\begin{equation}
\mathcal{I}(\tau)=\epsilon\int_0^\tau dt\int
d\textbf{r}\Omega(\textbf{r})
\tilde{\Psi}^\dag(\textbf{r},t)\tilde{\Psi}(\textbf{r},t).
\label{eq-intensity}
\end{equation}
This equation is formally equivalent to the quantum counting
formula, Eq. (\ref{eq-quantum-mandel}). However, whereas for the
quantum counting formalism the intensity of particles registered
at the detector is determined by the square of the field operator,
in Eq. (\ref{eq-intensity}) the intensity is calculated using a
modified field operator $\tilde{\Psi}(\textbf{r},t)$, which
includes the absorption at the detector. In the following, we
analyze these modified field operators.

We rewrite eq. (\ref{eq-def-psitilde}) by dividing the detection time into small sub-intervals $\Delta t=t/N$. The time integration in Eq. (\ref{eq-def-F(t)}) can be written as a sum, such
that
$F(t)=\prod_i F_i(t_i)$,
with $F_i(t_i)=e^{-\frac{\epsilon}{2}\Delta t\int
d\textbf{r}'\Omega(\textbf{r}')\Psi^\dag(\textbf{r}',t_i)\Psi(\textbf{r}',t_i)}$,
and we get
\begin{eqnarray}
\tilde{\Psi}(\textbf{r},t)=
F^{-1}_1(t_1)...F_N^{-1}(t_N)\Psi(\textbf{r},t)F_N(t_N)...F_1(t_1).
\end{eqnarray}

We evaluate the expressions by using the commutation relations
stated above. We start with the inner term,
\begin{equation}
F_N^{-1}(t_N)\Psi(\textbf{r},t)F_N(t_N)=e^{-\epsilon\Delta
t\Omega(\textbf{r})}\Psi(\textbf{r},t_N).\label{eq-first-term}
\end{equation}
The second term thus reads
\begin{eqnarray}
e^{-\epsilon\Delta
t\Omega(\textbf{r})}F_{N-1}^{-1}(t_{N-1})\Psi(\textbf{r},t_N)
F_{N-1}(t_{N-1})=\nonumber\\e^{-{\epsilon}\Delta t
\Omega(\textbf{r})}\int d\textbf{r}'e^{-{\epsilon}\Delta t
\Omega(\textbf{r}')}G_0(\textbf{r},\textbf{r}',t_N-t_{N-1})\Psi(\textbf{r}',t_{N-1})
\label{eq-second-term}
\end{eqnarray}
The successive terms are calculated analogously, and in the limit
of infinitesimal small time intervals we get
\begin{equation}
\tilde{\Psi}(\textbf{r},t)=\int
d\textbf{r}'G(\textbf{r},\textbf{r}',t,t_0)\Psi(\textbf{r}',t_0),\label{eq-Psi-tilde}
\end{equation}
where
\begin{equation}
G(\textbf{r},\textbf{r}',t,t_0)=\bra{\textbf{r}}e^{-i(t-t_0)/\hbar(\hat
H_0-i{\epsilon}\hbar\hat\Omega)}\ket{\textbf{r}'}.\label{eq-def-G}
\end{equation}
We have thus obtained an expression for the modified field
operators $\tilde{\Psi}(\textbf{r},t)$ which differs from the
definition of the standard field operator as it includes the
propagation in an imaginary potential created by the detector.
 Together with the counting formula Eq.~(\ref{eq-p-time-dependent}), this allows us, in principle, to
calculate the counting distribution for time dependent systems
with arbitrary detector geometries. However, solving Eq.~(\ref{eq-Psi-tilde}) is in general a highly non-trivial task. In Sect. \ref{sec-system-td} we solve the equation for the detection of an expanding BEC.

It is interesting to point out that in many experimental situations, the detection process is fast
compared to the time evolution of the system. In this case, we can
neglect the part corresponding to the Hamiltonian $H_0$ in eq.
(\ref{eq-Psi-tilde}) and get
\begin{eqnarray}
\tilde{\Psi}(\textbf{r},t)=\int
d\textbf{r}'\bra{\textbf{r}}e^{-{\epsilon}\hat\Omega(t-t_0)}\ket{\textbf{r}'}\Psi(\textbf{r}',t_0)\nonumber\\
=e^{-{\epsilon}\hat\Omega(t-t_0)}\Psi(\textbf{r},t_0).\label{eq-psi-tilde-exact}
\end{eqnarray}
The intensity (\ref{eq-intensity}) thus reads
\begin{eqnarray}
\mathcal{I}(\tau)=\epsilon\int
d\textbf{r}(1-e^{-\epsilon\Omega(\textbf{r})\tau})\Psi^\dag(\textbf{r},t_0)\Psi(\textbf{r},t_0),\label{eq-intensity-short-time}
\end{eqnarray}
which is a generalization of the formula Eq. (\ref{eq-mollow})
considering finite detector volumes. For $\epsilon\tau\ll 1$, Eq.
(\ref{eq-intensity-short-time}) reduces to the quantum counting
formula Eq. (\ref{eq-quantum-mandel}) for time independent
systems.

\section{Detection with Two Detectors}\label{sec:2d}
Our formalism is easily extended to calculate the joint counting
probability for the detection at two detectors. The master
equation that describes counting with two detectors reads
\begin{eqnarray}
&&\dot{\rho}(t)=\epsilon_1 \int d\textbf{r}
\Omega_1(\textbf{r})\Psi(\textbf{r},t)\rho
\Psi^\dag(\textbf{r},t)\nonumber\\
&&+\epsilon_2 \int d\textbf{r}
\Omega_2(\textbf{r})\Psi(\textbf{r},t)\rho
\Psi^\dag(\textbf{r},t)\nonumber\\&&-\frac{\epsilon_1}{2}\int
d\textbf{r}\Omega_1(\textbf{r})(\Psi(\textbf{r},t)^\dag
\Psi(\textbf{r},t) \rho- \rho \Psi(\textbf{r},t)^\dag
\Psi(\textbf{r},t))\nonumber\\
&&-\frac{\epsilon_2}{2}\int
d\textbf{r} \Omega_2(\textbf{r})(\Psi(\textbf{r},t)^\dag
\Psi(\textbf{r},t) \rho-\rho \Psi(\textbf{r},t)^\dag
\Psi(\textbf{r},t)),\label{eq-master-Psi-2}
\end{eqnarray}
Similarly to the case of one detector described in Sect. \ref{sec-time-dep}, we solve the master equation eq. (\ref{eq-master-Psi-2}) by performing the transformation
$\rho(t)=F_2(t)\tilde{\rho}(t)F_2^\dag(t)$ and
\begin{equation}
\tilde{\Psi}(\textbf{r},t)=F_2^{-1}(t)\Psi(\textbf{r},t)F_2(t),\label{eq-def-psitilde_2}
\end{equation}
where the operator $F_2(t)$ is defined as
\begin{widetext}
\begin{eqnarray}
F_2(t)=\mathcal{T}e^{-\epsilon_1/2\int_0^t dt' \int
d\textbf{r}'\Omega_1(\textbf{r}')\Psi^\dag(\textbf{r}',t')\Psi(\textbf{r}',t')-\epsilon_2/2\int_0^t dt' \int
d\textbf{r}'\Omega_2(\textbf{r}')\Psi^\dag(\textbf{r}',t')\Psi(\textbf{r}',t')}.
\label{eq-def-F_2(t)}
\end{eqnarray}
\end{widetext}
The evolution of the rotated density matrix is thus given by
\begin{eqnarray}
\dot{\tilde{\rho}}(\textbf{r},t)=\epsilon_1\int
d\textbf{r}'\Omega_1(\textbf{r}')\tilde{\Psi}(\textbf{r},t)\tilde{\rho}\tilde{\Psi}^\dag(\textbf{r},t)\nonumber\\+\epsilon_2\int
d\textbf{r}'\Omega_2(\textbf{r}')\tilde{\Psi}(\textbf{r},t)\tilde{\rho}\tilde{\Psi}^\dag(\textbf{r},t),\label{eq-dot_rhotilde_2}
\end{eqnarray}
The equation can be solved using perturbation theory, where we get
an expression as in eq. (\ref{eq-rhotilde}) that includes
correlation terms between the two detectors.

The conditional probability distribution of counting $m$ particles
in one detector and $n$ particles in the other one thus reads
\begin{eqnarray}
p(m,n)=\frac{(-1)^{m+n}}{m!n!}\frac{d^{m+n}}{d\lambda_1^md\lambda_2^n}\mathcal{Q}\Big|_{\lambda_1=1,\lambda_2=1},\label{joined_p}
\end{eqnarray}
where
\begin{equation}
\mathcal{Q}(\lambda_1,\lambda_2)=\mbox{Tr}(\rho\mathcal{T}:e^{-\lambda_1\mathcal{I}_1-\lambda_2\mathcal{I}_2}:),
\end{equation}
and
\begin{equation}
\mathcal{I}_i(\tau)=\epsilon_i\int_0^\tau dt\int
d\textbf{r}\Omega_i(\textbf{r})
\tilde{\Psi}^\dag(\textbf{r},t)\tilde{\Psi}(\textbf{r},t).
\label{eq-intensity}
\end{equation}
The modified field operator that includes the absorption at the
two detectors is given by
\begin{equation}
\tilde{\Psi}(\textbf{r},t)=\int
d\textbf{r}'\bra{\textbf{r}}e^{-i(t-t_0)/\hbar(\hat
H_0-i\epsilon\hbar(\hat\Omega_1+\hat\Omega_2))}\ket{\textbf{r}'}\Psi(\textbf{r}',t_0).\label{eq-Psi-tilde-2det}
\end{equation}

\section{Detection of an Expanding Bose-Einstein Condensate}\label{sec-system-td}
Let us now illustrate the space and time dependent counting
process by considering the counting statistics of a freely
expanding Bose Einstein condensate.  For simplicity, we consider a
one dimensional system with a point-like detector located at a
distance $z_0$ from the condensate. The detection time is of the
order of the system dynamics, such that we calculate the full
time- and space dependent generating function with the intensity
given by eq. (\ref{eq-intensity}). We consider a point-like
detector placed at a distance $z_0$ from the center of the cloud.
The detector is modeled by a delta-function $\delta(z-z_0)$, such
that the intensity eq. (\ref{eq-intensity}) reads
\begin{equation}
\mathcal{I}=\epsilon\int_0^tdt'\tilde{\Psi}^\dag(z_0,t')\tilde{\Psi}(z_0,t')\label{eq-intensity-delta},
\end{equation}
where the time evolution of the operators
$\tilde{\Psi}(\textbf{r},t)$ is described by eq.
(\ref{eq-Psi-tilde}). For the detection of a 1D BEC at a
point-like detector, the time evolution of the single-particle
wave function is given by
\begin{equation}
\tilde{\phi}(z,t)=\int
dz'G(z,z',t,t_0)\phi(z',t_0),\label{eq-Psi-tilde-td}
\end{equation}
where \begin{equation}
G(z,z',t,t_0)=\bra{z}e^{-i(t-t_0)/\hbar(\hat
H_0-i\hbar\epsilon\hat\delta)}\ket{z'}\label{eq-GS}
\end{equation}

The counting distribution is then obtained from eq.
(\ref{eq-p-time-dependent}), which for the case of a condensate
with $N$ particles reads
\begin{equation}
p(m)=\frac{\left(N \epsilon\int dt
\tilde{\phi}^\dag(z_0,t)\tilde{\phi}(z_0,t) \right)^m}{m!}e^{-N
\epsilon\int dt
\tilde{\phi}^\dag(z_0,t)\tilde{\phi}(z_0,t)},\label{eq-p-td}
\end{equation}
In the following subsection, we exactly solve eq.
(\ref{eq-Psi-tilde-td}), where we approximate the initial wave
function $\tilde{\phi}(z,0)$ by a Lorentzian function,
\begin{equation}
 \tilde{\phi}(z,0)=\sqrt{\Gamma}e^{-\Gamma |z|}\label{eq-lorentz}.
\end{equation}
In Subsect. \ref{sec-born}, we calculate an approximate solution
obtained by the Born approximation. In Subsect.
\ref{sec-results-td} we compare the exact solution to the
approximate solution as well as to the quantum counting formalism.
\subsection{Exact solution}
\label{sec-propagater-td}
\begin{figure}
\begin{center}
\epsfig{file=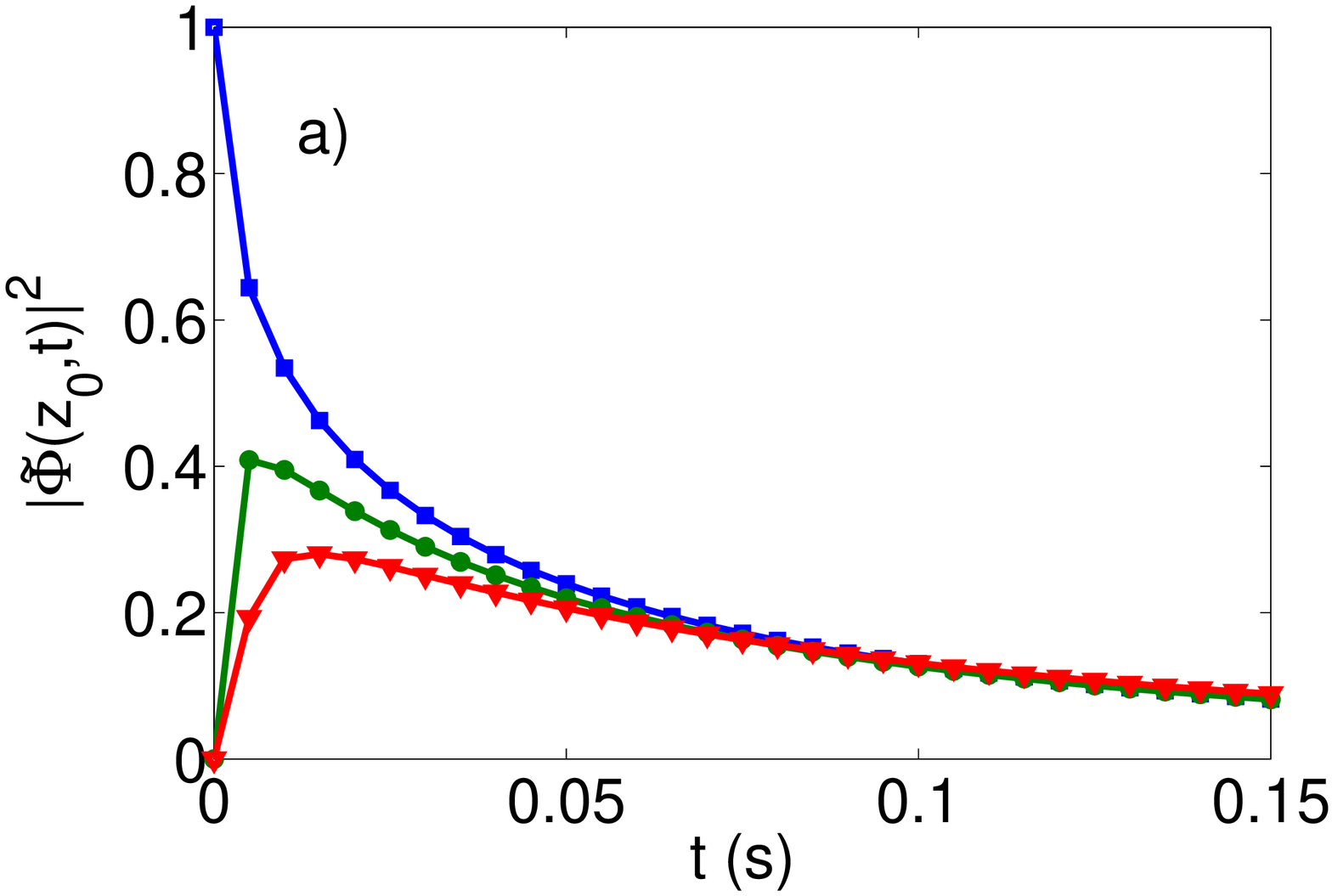 ,width=0.45\linewidth}\epsfig{file=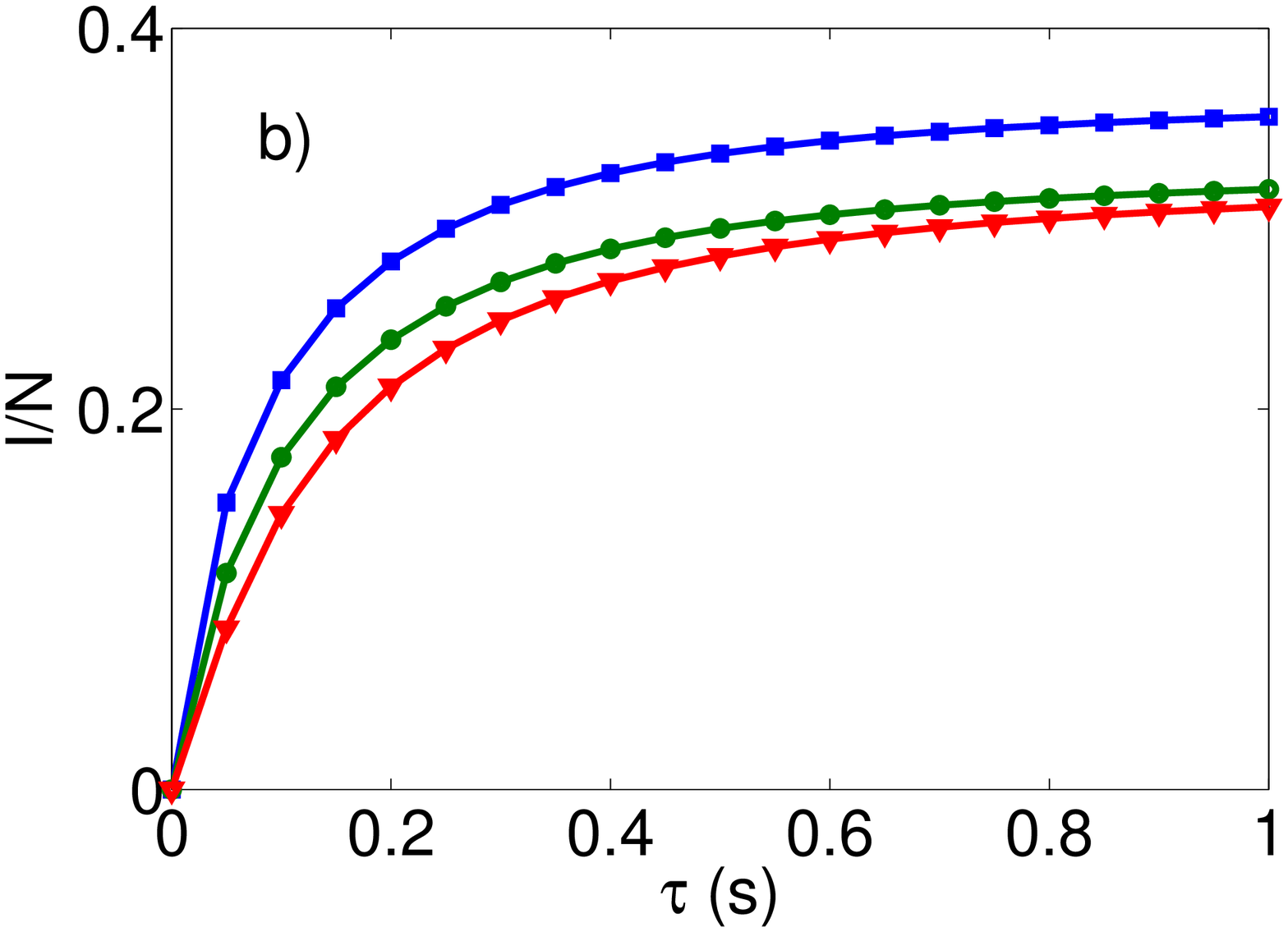
,width=0.45\linewidth}\caption{Density $|\tilde{\phi}(z_0,t)|^2$
with respect to time (Fig a) and normalized intensity
$\mathcal{I}/N$ with respect to $\tau$ (Fig. b) for $z_0=0$ (blue
squares), $z_0=0.1/\Gamma$ (green circles), $z_0=0.4/\Gamma$ (red
diamonds). As the distance $z_0$ between the center of the cloud
and the detector increases, the wave function spreads and long
detector opening times are required to achieve non-trivial
intensities. Parameters used: $z_0=0, \epsilon=1, N=100,
\Gamma=10^5$.} \label{fig-densityZ}
\end{center}
\end{figure}

We calculate the counting distribution
$p(m)$ for an expanding BEC, where the detector is located at some
distance $z_0$ from the center of the cloud. We follow the
treatment in
 \cite{Kleber1994} to derive an exact solution for the
propagator eq. (\ref{eq-GS}) that describes the whole system
evolution including the absorption at the detector. The system
Hamiltonian is composed of two parts: the free particle
Hamiltonian $H_0$ and a part corresponding to the detection
process, which acts as an imaginary potential. The propagator
$G(z,z',t)$ that describes the time evolution of the wave-function
including the absorption at the detector can be written in an
iterative way using the Lippmann-Schwinger equation, which for a
point-like detector $\Omega(z')=\delta(z'-z_0)$ reads
\begin{eqnarray}
G(z,z',t)=G_0(z,z',t)-\nonumber\\\frac{\epsilon}{\hbar}\int_0^t
dt'G_0(z,z_0,t-t')G(z_0,z',t').\label{eq-G-LS-delta}
\end{eqnarray}
In the Appendix, we solve the Lippmann-Schwinger equation and show
that the full propagator that describes the back-action of the
detector is given by
\begin{eqnarray}
G(z_0,z',t)=G_0(z_0,z't)\nonumber\\+\frac{i m
\epsilon}{\hbar}M(|z-z_0|+|z'-z_0|,-\frac{m\epsilon}{\hbar},\frac{\hbar}{m}t),\label{eq-G-exact}
\end{eqnarray}
where $M(z,k,t)$ denotes the Moshinsky function \cite{Kramer2005}
\begin{eqnarray}
M(z,k,t)=\int_{-\infty}^0dz'\frac{1}{\sqrt{2\pi i
t}}\exp\left(ikz'+i\frac{(z-z')^2}{2t}\right)\nonumber\\
=\frac{1}{2}\exp\left(ikz-i\frac{1}{2}k^2t\right)\hbox{erfc}(e^{-i\pi/4}\frac{z-kt}{\sqrt{2t}})\label{eq-Moshinsky}
\end{eqnarray}
The modified wave function $\tilde{\phi}(z_0,t)$ can then be
calculated by standard integration techniques using eq.
(\ref{eq-Psi-tilde-td}), and the counting distribution $p(m)$
obtained by eq. (\ref{eq-p-td}). The counting statistics are
determined by the time integral over the square of the wave
function $\tilde{\phi}(z_0,t)$. Fig. \ref{fig-densityZ} a shows
the square of the wave function
 with respect to time
for different distances $z_0$. As the distance increases, the wave
function spreads in time. In order to obtain non-trivial results,
long opening times are required. Fig. \ref{fig-densityZ} b shows
the intensity of particles at the detector with respect to the
opening time $\tau$ for detectors placed at different distances
from the detector.

\subsection{Born Approximation} \label{sec-born}In the previous
section, we obtained an exact solution for the time evolution of
the wavefunction by solving the Lippmann-Schwinger equation
(\ref{eq-G-LS-delta}). In this section, we use the Born
approximation in order to derive an expression for
$\tilde{\phi}(z,t)$ in terms of the known propagator $G_0$. In the
second order approximation, we obtain
\begin{equation}
G(z,z',t)=G_0(z,z',t)-\frac{\epsilon}{\hbar}\int_0^t
dt'G_0(z,z_0,t-t')G_0(z_0,z',t').\label{eq-G-born}
\end{equation}
This implies that up to second order, the solution to eq.
(\ref{eq-Psi-tilde-td}) is given by
\begin{eqnarray}
&&\tilde{\phi}(z,t)=\phi(z,t)\nonumber\\&&-\frac{\epsilon}{\hbar}\int_0^t
dt'\int dz'G_0(z,z_0,t-t')G_0(z_0,z',t')
\phi(z',t_0).\nonumber\\
&&=2\sqrt{\Gamma}(M(z_0,-i\Gamma,\hbar
t/m)-\nonumber\\&&\frac{\epsilon}{\hbar} \int_0^t
dt'\sqrt{\frac{m}{2\pi \hbar i t'}} M(z_0,-i\Gamma,\hbar t'/m))
\label{eq-Psi-tilde-first-order}
\end{eqnarray}
Eq. (\ref{eq-Psi-tilde-first-order}) describes the evolution of
the wave function, where the absorption at the detector is taken
into account up to second order. We get the higher order Born
approximation by writing eq. (\ref{eq-Psi-tilde-first-order}) in
exponential form,
\begin{eqnarray}
\tilde{\phi}(\textbf{z},t)=\sqrt{\Gamma}M(z_0,-i\Gamma,\hbar t/m)
\times\nonumber\\e^{-\epsilon/\hbar\int_0^t
dt'\sqrt{\frac{m}{2\pi\hbar i t'}} M(z_0,-i\Gamma,\hbar
t'/m)/M(z_0,-i\Gamma,\hbar t/m)}\label{eq-born-exponential}
\end{eqnarray}

In Sect. \ref{sec-results-td}, we show that the Born approximation
describes the situation more accurately than the quantum counting
formalism. However, the effect of the absorption is
under-estimated.

\subsection{Effect of the Absorption at the Detector}\label{sec-results-td} Let us now analyze the effect of
the back-action of the detector on the field. From eq.
(\ref{eq-p-td}) it is clear that the important quantities to study
are the square of the wave function,
$\tilde{\phi}^*(z_0,t)\tilde{\phi}(z_0,t)$, its time integral, as
well as the full counting distribution. We discuss the limits in
which the quantum counting formalism and the Born approximation
give valid results, and study the limitations of the approximative
solutions.

\begin{figure}
\epsfig{file=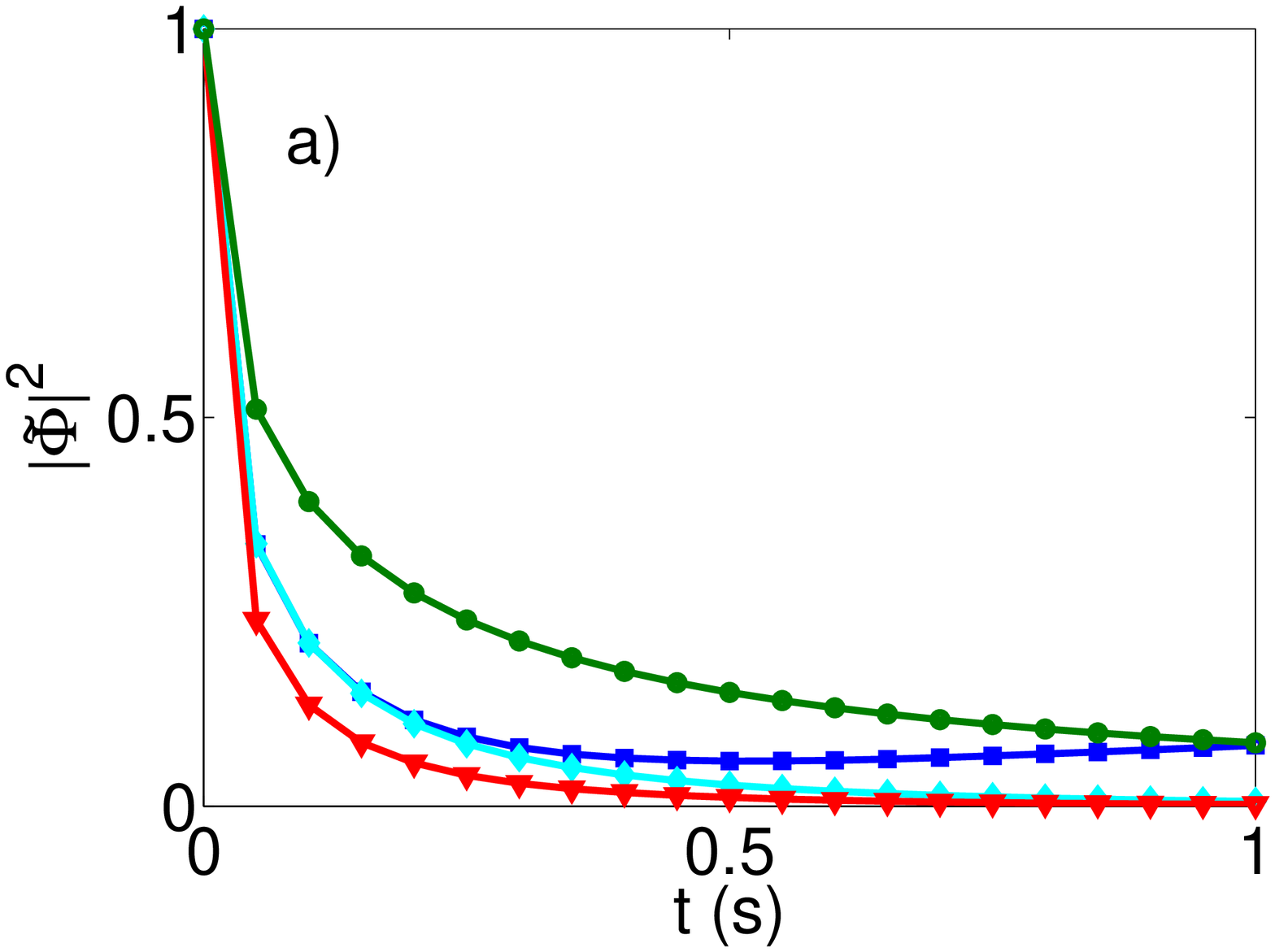 ,width=0.48\linewidth} \epsfig{file=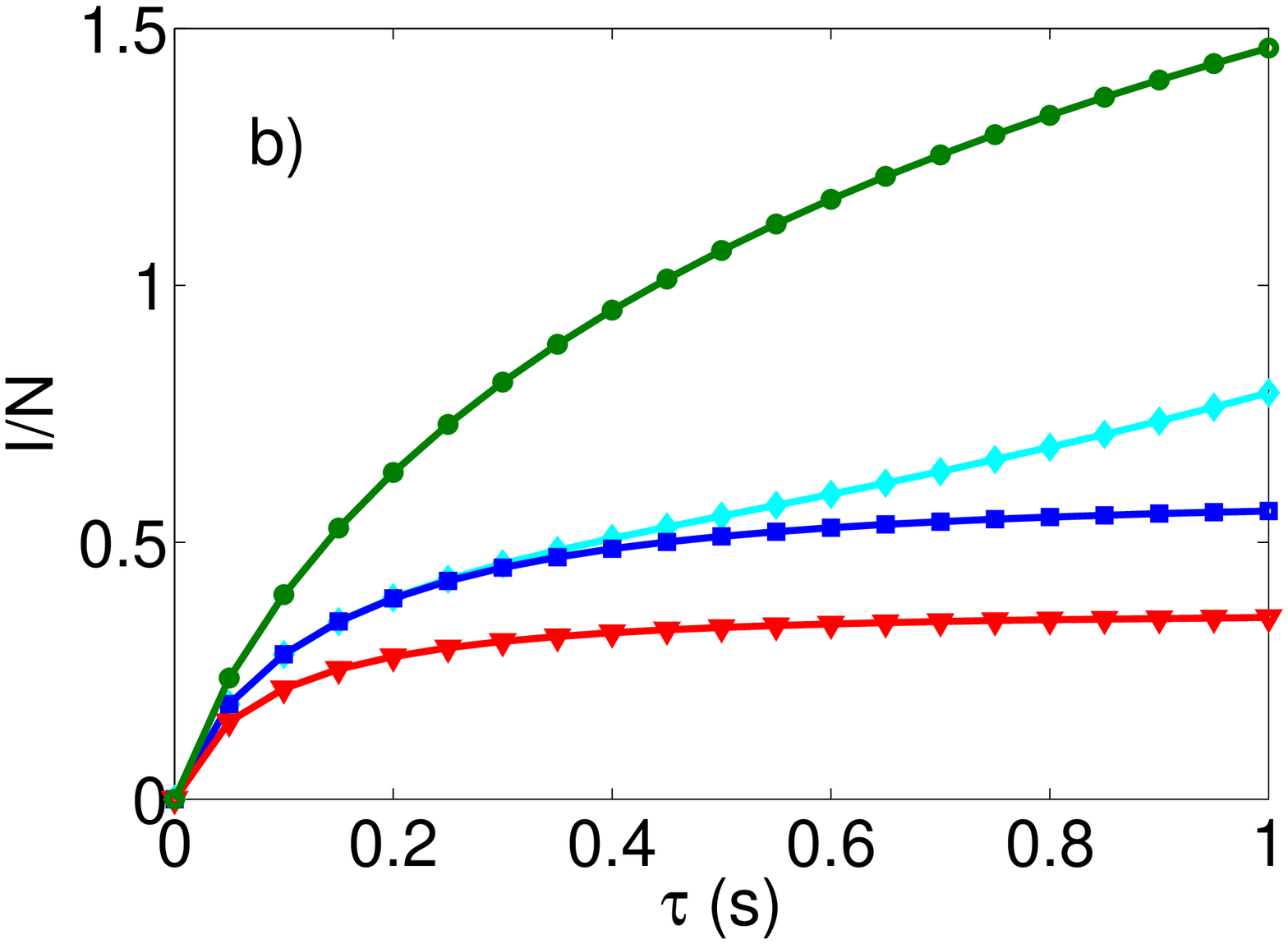
,width=0.48\linewidth}\caption{a) $|\tilde{\phi}(0,t)|^2$ with
respect to $t$. b) Normalized intensity $\mathcal{I}/N$ with
respect to $\tau$. We compare the exact solution (red triangles),
second order Born approximation (blue squares), exponential Born
approximation (light blue diamonds) and the quantum counting
formalism (green circles). Both the quantum counting formalism and
the Born approximation reach intensities that exceed the total
number of particles for long opening times $\tau$. In the
exponential Born approximation the intensity is bounded for large
$\tau$, however, the asymptotic value exceeds the bound given by
the exact solution. Parameters used: $z_0=0, \epsilon=1, N=100,
\Gamma=10^5$.} \label{fig-psi-e-m-b}
\end{figure}

In Fig. \ref{fig-psi-e-m-b}a, we plot the square of the wave
function with respect to time, and compare the exact solution to
the solutions obtained by the born approximation and the quantum
counting formalism. The exact graph corresponding to the exact
solution decays more rapidly, as the absorption at the detector is
considered. The Born approximation underestimates the decay of the
wave function and thus the absorption, however, it describes the
behavior more accurately than the quantum counting formalism,
where absorption is not considered.

The effect is seen more clearly when studying the intensity of the
field at the detector, which is given by the time integral
$\epsilon\int_0^\tau dt \tilde{\phi}^*(z_0,t)\tilde{\phi}(z_0,t)$
(Fig. \ref{fig-psi-e-m-b}b). For short detection times, the exact
solution and the approximate solutions coincide. As the detection
time increases, the intensity of particles is overestimated both
for the Born approximation and for the quantum counting formalism.
Note that for long detection times, the second order Born
approximation diverges, whereas the exponential Born approximation
reaches an asymptotic value.

\begin{figure}
\epsfig{file=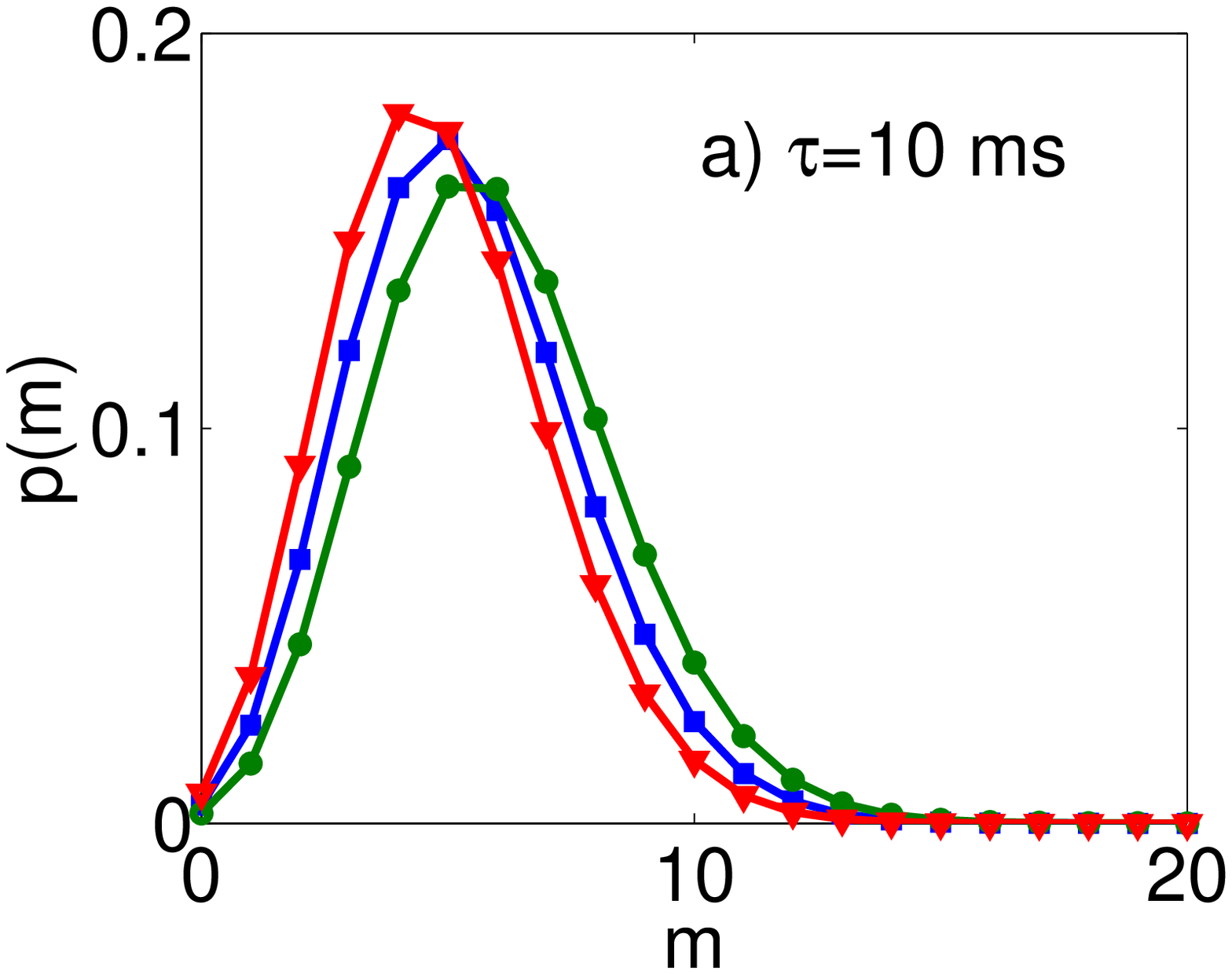 ,width=0.45\linewidth}
\epsfig{file=
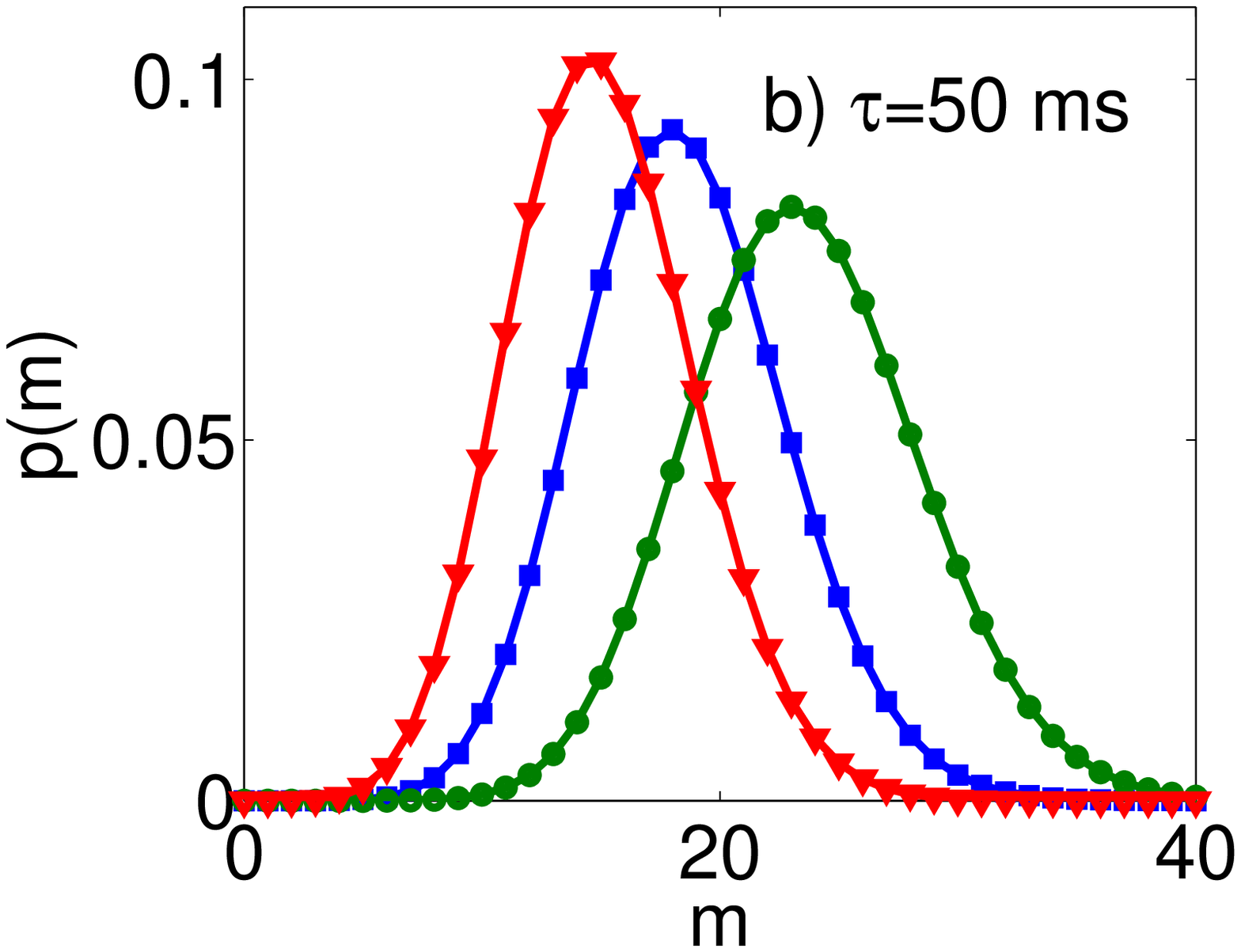,width=0.45\linewidth}\caption{Counting
distribution p(m) obtained from exact solution (red triangles),
Born approximation (blue squares) and the quantum counting
formalism (green circles). For short detection times (Fig. a) the
approximations agree reasonably well with the exact solution. For
longer detection times (Fig. b), the approximations are no longer
valid. Parameters used: $z_0=0, \epsilon=1, N=100, \Gamma=10^5$.}
\label{fig-distributions-td}
\end{figure}

Finally, we compare the counting distributions obtained by the
exact solution eq. (\ref{eq-Psi-tilde-td}) to the solution
obtained by the quantum counting formalism and the Born
approximation. The effect of absorption is clearly visible in the
counting distribution, where the approximate solutions deviate
increasingly from the exact solution as the measurement time
increases (Fig. \ref{fig-distributions-td}). The counting
distribution calculated with the full formalism including
detection is clearly different from the approximated solution.

\section{Summary and Conclusions}\label{sec:summ}
We have derived a formalism for describing the counting
distribution of space- and time dependent fields taking into
account the back-action of the detector on the field. We have
illustrated the importance of the effect of the back-action for
the free expansion of a Bose Einstein condensate. An approximate
solution using the Born approximation describes the behavior of
the system more accurately than the quantum counting formalism.
However, for typical detection times of expanding BECs, the effect
of absorption is under-estimated significantly both by the quantum
counting formalism, as well as the Born approximation. We thus
showed that for certain experimentally relevant situations, the
full time- and space dependent formalism has to be applied.
\bibliographystyle{unsrtnat}
\begin{acknowledgments}
We acknowledge financial support from the Spanish MICINN project
FIS2008-00784 (TOQATA), FIS2010-18799, Consolider Ingenio 2010
QOIT, EU-IP Project AQUTE, EU STREP project NAMEQUAM, ERC Advanced
Grant QUAGATUA, the Ministry of Education of the Generalitat de
Catalunya. M.R. is grateful to the MICINN of Spain for a Ram\'on y
Cajal grant, M.L. acknowledges the Alexander von Humboldt
Foundation and Hamburg Theoretical Physics Prize.
\end{acknowledgments}
\appendix*
\section{Derivation of the Propagator Including the Back-action of
the Detector} \label{Ap-z0} We solve the Lippmann-Schwinger
equation for the propagator $G(z,z',t)$ that describes the time
evolution of the wave-function including the absorption at a
point-like detector $\Omega(z')=\delta(z'-z_0)$,
\begin{eqnarray}
G(z,z',t)=G_0(z,z',t)-\nonumber\\\frac{\epsilon}{\hbar}\int_0^t
dt'G_0(z,z_0,t-t')G(z_0,z',t'),\label{eq-G-LS-delta-ap}
\end{eqnarray}
where the propagator $G_0(z,z',t)$ for the free expansion is given
by
\begin{equation}
G_0(z,z',t)=\sqrt{\frac{m}{2\pi i\hbar
t}}\exp(\frac{i|z-z'|^2}{2\hbar t}).
\end{equation}
 We perform a Laplace transform of
eq. (\ref{eq-G-LS-delta-ap}) and use the convolution theorem to
get
\begin{eqnarray}
\tilde{G}(z,z',s)=\tilde{G}_0(z,z',t)-\frac{\epsilon}{\hbar}\tilde{G}_0(z,z_0,s)\tilde{G}(z_0,z',s)\label{eq-Laplace-G}
\end{eqnarray}
From eq. (\ref{eq-intensity-delta}), we observe that we are only
interested in the propagator at $z=z_0$, such that
\begin{equation}
\tilde{G}(z_0,z',s)=\frac{\tilde{G}_0(z_0,z',t)}{1-\frac{\epsilon}{\hbar}\tilde{G}_0(z_0,z_0,s)}.\label{eq-Laplace-G-x0}
\end{equation}
The Laplace transform of the free propagator $G(z_0,z',s)$ is
given by
\begin{equation}
\tilde{G}(z_0,z',s)=\sqrt{\frac{m}{2i \hbar
s}}\exp(-\sqrt{\frac{2ms}{i\hbar}}|z_0-z'|).\label{eq-Laplace-G0}
\end{equation}
The inverse Laplace transform of $\tilde{G}(z,z',s)$ can be
performed by standard methods \cite{Abramowich1965}, such that the
propagator is given by
\begin{eqnarray}
G(z_0,z',t)=G_0(z_0,z't)\nonumber\\+\frac{i m
\epsilon}{\hbar}M(|z-z_0|+|z'-z_0|,-\frac{m\epsilon}{\hbar},\frac{\hbar}{m}t),\label{eq-G-exact}
\end{eqnarray}
where $M(z,k,t)$ is defined in eq. (\ref{eq-Moshinsky}). The
function $\tilde{\phi}(z_0,t)$ is then calculated by eq.
(\ref{eq-Psi-tilde-td}).

 Note that for a detector placed at
$z_0=0$ in the center of the cloud, the counting formula can also
be obtained by directly solving the time dependent Schr\"odinger
equation,
\begin{equation}
i\hbar\dot{\tilde{\phi}}(z,t)=-\frac{\hbar^2}{2m}\frac{\partial^2}{\partial
z^2}\tilde{\phi}(z,t)-i\hbar\epsilon\delta(z)\tilde{\phi}(z,t)
\label{eq-psi-ap}
\end{equation}
with the initial condition given by
\begin{equation*}
 \tilde{\phi}(z,0)=\sqrt{\Gamma}e^{-\Gamma |z|}.
\end{equation*}

In order to solve eq. (\ref{eq-psi-ap}), we first express it in
terms of its fourier transform. From eq.
(\ref{eq-intensity-delta}) is is clear that for a detector placed
at $z_0=0$, we are only interested in
$\tilde{\phi}(0,t)=\frac{1}{\sqrt{2\pi\hbar}}\int_{-\infty}^\infty
dk \phi(k,t)$. The fourier transformed equation is a differential
equation with variable coefficients and can be integrated by
standard methods \cite{Bronstein2002}. We get
\begin{equation}
\tilde{\phi}(0,t)=\tilde{\phi}_0(t)-\frac{\epsilon}{\sqrt{2\pi\hbar}}\int_0^t\kappa(t-t')\tilde{\phi}(0,t')dt',\label{eq-S(t)}
\end{equation}
where $\tilde{\phi}_0(t)=\frac{1}{\sqrt{2\pi\hbar}}\int dk
e^{\frac{-ik^2t}{2\hbar m}}\phi(k,0)$ and $
\kappa(t,t')=\frac{1}{\sqrt{2\pi\hbar}}\int_{-\infty}^\infty dk
e^{\frac{-ik^2(t-t')}{2\hbar
m}}=\frac{(1-i)\sqrt{m}}{\sqrt{2(t-t')}}$ We take the Laplace
transform of eq. (\ref{eq-S(t)}) and use the convolution theorem
\cite[29.2.8]{Abramowich1965} to obtain
\begin{equation}
\tilde{\phi}(0,s)=\frac{\tilde{\phi}_0(s)}{1+\frac{\epsilon}{\sqrt{2\pi\hbar}}
\kappa(s)}.\label{eq-S(s)}
\end{equation}
The term $\tilde{\phi}_0(s)$ is calculated by using the residue
theorem. Using the method of partial fractions the expression for
$\tilde{\phi}(0,s)$ is written in the form
$\frac{A}{\sqrt{s}+a_1}$, such that the inverse Laplace transform
is given by
\begin{widetext}
\begin{equation}
\tilde{\phi}(0,t)=\frac{\sqrt{\Gamma}(\hbar\Gamma
e^{\frac{i\hbar\Gamma^2 t}{2m}}\hbox{Erfc}(\frac{1+i}{2}
\sqrt{\hbar t/m}\Gamma)+i m\epsilon e^{-\frac{im t
\epsilon^2}{2\hbar}}
\hbox{Erfc}(\frac{1-i}{2}\epsilon\sqrt{tm/\hbar})}{(\hbar\Gamma+im\epsilon)}.
\end{equation}
\end{widetext}

\bibliography{references}

\begin{thebibliography}{29}
\providecommand{\natexlab}[1]{#1}
\providecommand{\url}[1]{\texttt{#1}}
\expandafter\ifx\csname urlstyle\endcsname\relax
  \providecommand{\doi}[1]{doi: #1}\else
  \providecommand{\doi}{doi: \begingroup \urlstyle{rm}\Url}\fi

\bibitem[Mandel(1958)]{Mandel1958}
L.~Mandel.
\newblock Fluctuations of photon beams and their correlations.
\newblock \emph{Proc. Phys. Soc (London)}, 72:\penalty0 1037, 1958.

\bibitem[Mandel(1959)]{Mandel1959}
L.~Mandel.
\newblock Fluctuations of photon beams - the distribution of the photo
  electrons.
\newblock \emph{Proc. Phys. Soc (London)}, 74:\penalty0 233, 1959.

\bibitem[Mandel(1963)]{Mandel1963}
L.~Mandel.
\newblock Fluctuations of light beams.
\newblock \emph{in: Progress in Optics}, page 181, 1963.
\newblock ed. E. Wolf (North-Holland, Amsterdam).

\bibitem[Glauber(1963{\natexlab{a}})]{Glauber1963a}
R.~J. Glauber.
\newblock Photon correlations.
\newblock \emph{Phys. Rev. Lett.}, 10:\penalty0 84, 1963{\natexlab{a}}.

\bibitem[Glauber(1963{\natexlab{b}})]{Glauber1963b}
R.~J. Glauber.
\newblock Coherent and incoherent states of the radiation field.
\newblock \emph{Phys. Rev.}, 131:\penalty0 2766--2788, 1963{\natexlab{b}}.

\bibitem[Glauber(1963{\natexlab{c}})]{Glauber1963c}
R.~J. Glauber.
\newblock The quantum theory of optical coherence.
\newblock \emph{Phys. Rev.}, 130:\penalty0 2529--2539, 1963{\natexlab{c}}.

\bibitem[Glauber(1965)]{Glauber1965}
R.~J. Glauber.
\newblock Optical coherence and photon statistics.
\newblock page~65, New York, 1965. Gordon \& Breach.

\bibitem[Kelley and Kleiner(1964)]{Kelley1964}
P.~L. Kelley and W.~H. Kleiner.
\newblock Theory of electromagnetic field measurement and photoelectron
  counting.
\newblock \emph{Phys. Rev.}, 136\penalty0 (A316-A334), 1964.

\bibitem[Mollow(1968)]{Mollow1968}
B.~R. Mollow.
\newblock Quantum theory of field attenuation.
\newblock \emph{Phys. Rev.}, 168\penalty0 (1896), 1968.

\bibitem[Scully and W.~E.~Lamb(1969)]{Scully1969}
M.~O. Scully and Jr. W.~E.~Lamb.
\newblock Quantum theory of an optical maser. iii. theory of photoelectron
  counting statistics.
\newblock \emph{Phys. Rev.}, 179\penalty0 (368), 1969.

\bibitem[Selloni et~al.(1978)Selloni, Schwendimann, Quattropani, and
  Baltes]{Selloni1978}
A.~Selloni, P.~Schwendimann, P.~Quattropani, and H.~P. Baltes.
\newblock Open system theory of photodetection: dynamics of field and atomic
  moments.
\newblock \emph{J. Phys. A}, 11\penalty0 (1427), 1978.

\bibitem[Srinivas and Davies(1981)]{Srinivas1981}
M.~D. Srinivas and E.~B. Davies.
\newblock Photon counting probabilities in quantum optics.
\newblock \emph{Opt. Acta}, 28:\penalty0 981--996, 1981.

\bibitem[Imoto et~al.(1990)Imoto, Ueda, and Ogawa]{Imoto1990}
N.~Imoto, M.~Ueda, and T.~Ogawa.
\newblock Microscopic theory of continuous measurement of photon number.
\newblock \emph{Phys. Rev. A}, 41\penalty0 (7):\penalty0 4127--4129, 1990.

\bibitem[Dodonov et~al.(2007)Dodonov, Mizrahi, and Dodonov]{Dodonov2007}
A.~V. Dodonov, S.~S. Mizrahi, and V.~V. Dodonov.
\newblock Inclusion of nonidealities in the continuous photodetection.
\newblock \emph{Phys. Rev. A}, 75\penalty0 (013806), 2007.

\bibitem[H\"ayrynen et~al.(2010)H\"ayrynen, Oksanen, and Tulkki]{Hayrynen2010a}
T.~H\"ayrynen, J.~Oksanen, and J.~Tulkki.
\newblock Derivation of generalized quantum jump operators and comparison of
  the microscopic single photon detector models.
\newblock \emph{Eur. Phys. J. D}, 56:\penalty0 113--121, 2010.

\bibitem[Ueda et~al.(1990)Ueda, Imoto, and Ogawa]{Ueda1990}
M.~Ueda, N.~Imoto, and T.~Ogawa.
\newblock Quantum theory for continuous photodetection processes.
\newblock \emph{Phys. Rev. A}, 41\penalty0 (7):\penalty0 3891--3904, 1990.

\bibitem[Parigi et~al.(2007)Parigi, Zavatta, Kim, and Bellini]{Parigi2007}
V.~Parigi, A.~Zavatta, M.~Kim, and M.~Bellini.
\newblock Probing quantum commutation rules by addition and subtraction of
  single photons to/from a light field.
\newblock \emph{Science}, 317\penalty0 (1890-1893), 2007.

\bibitem[H\"ayrynen et~al.(2009)H\"ayrynen, Oksanen, and Tulkki]{Hayrynen2009}
T.~H\"ayrynen, J.~Oksanen, and J.~Tulkki.
\newblock Exact theory for photon subtraction for fields from quantum to
  classical limit.
\newblock \emph{Eur. Phys. Lett.}, 78\penalty0 (44002), 2009.

\bibitem[Mandel(1981)]{Mandel1981}
L.~Mandel.
\newblock Comment on 'photon counting probabilities in quantum optics'.
\newblock \emph{Opt. Acta}, 28:\penalty0 1447--1450, 1981.

\bibitem[Fleischhauer and Welsch(1991)]{Fleischhauer1991}
M.~Fleischhauer and D.~G. Welsch.
\newblock Nonperturbative approach to multimode photodetection.
\newblock \emph{Phys. Rev. A}, 44\penalty0 (1):\penalty0 747--755, 1991.

\bibitem[Chmara(1987)]{Chmara1987}
W.~Chmara.
\newblock A quantum open-systems theory approach to photodetection.
\newblock \emph{J. Mod. Opt.}, 34\penalty0 (455-467), 1987.

\bibitem[Braungardt et~al.(2008)Braungardt, Sen, Sen, Glauber, and
  Lewenstein]{Braungardt2008}
S.~Braungardt, A.~Sen, U.~Sen, R.~J. Glauber, and M.~Lewenstein.
\newblock Fermion and spin counting in strongly correlated systems.
\newblock \emph{Phys. Rev. A}, 78:\penalty0 063613, 2008.

\bibitem[Braungardt et~al.(2011{\natexlab{a}})Braungardt, Rodríguez, Sen(De),
  Sen, Glauber, and Lewenstein]{Braungardt2011a}
S.~Braungardt, M.~Rodríguez, A.~Sen(De), U.~Sen, R.~J. Glauber, and
  M.~Lewenstein.
\newblock Counting of fermions and spins in strongly correlated systems in and
  out of thermal equilibrium.
\newblock \emph{Phys. Rev. A}, 83:\penalty0 013601, 2011{\natexlab{a}}.

\bibitem[Braungardt et~al.(2011{\natexlab{b}})Braungardt, Rodríguez, Sen(De),
  Sen, and Lewenstein]{Braungardt2011b}
S.~Braungardt, M.~Rodríguez, A.~Sen(De), U.~Sen, and M.~Lewenstein.
\newblock Atom counting in expanding ultracold clouds.
\newblock \emph{arXiv:1103.1868v1 [quant-ph]}, 2011{\natexlab{b}}.

\bibitem[Mandel and Wolf(1995)]{Mandel1995}
L.~Mandel and E.~Wolf.
\newblock \emph{Optical Coherence and quantum optics}.
\newblock Cambridge University Press, Cambridge, 1995.

\bibitem[Kleber(1994)]{Kleber1994}
M.~Kleber.
\newblock Exact solutions for time-dependent phenomena in quantum mechanics.
\newblock \emph{Physics Reports}, 236:\penalty0 331--393, 1994.

\bibitem[Abramowitz and Stegun(1965)]{Abramowich1965}
M.~Abramowitz and I.A. Stegun.
\newblock \emph{Handbook of Mathematical Functions}.
\newblock Dover, New York, 1965.

\bibitem[Kramer and Moshinsky(2005)]{Kramer2005}
T.~Kramer and M.~Moshinsky.
\newblock Tunnelling out of a time-dependent well.
\newblock \emph{J. Phys. A}, 38:\penalty0 5993--6003, 2005.

\bibitem[Bronstein et~al.(2002)Bronstein, Semendjajew, Musiol, and
  Muehlig]{Bronstein2002}
I.~N. Bronstein, K.~A. Semendjajew, G.~Musiol, and H.~Muehlig.
\newblock \emph{Handbook of Mathematics}.
\newblock Springer, 2002.

\end{thebibliography}

\end{document}